\documentclass[amsmath,amssymb,nofootinbib,10pt,eqsecnum,showkeys,showpacs,aps,prd]{elsarticle}
\usepackage{graphicx,amsmath,amsmath,amssymb}

\usepackage[breaklinks,colorlinks]{hyperref}
\usepackage{hyperref}
\hypersetup{%
    ,urlcolor=black
    ,citecolor=black
    ,linkcolor=black
    }

\def\be{\begin{equation}}
\def\ee{\end{equation}}
\def\bea{\begin{eqnarray}}
\def\eea{\end{eqnarray}}
\newcommand{\sech}[0]{\textrm{ sech}}

\begin{document}

\title{Photon interactions with superconducting topological defects}
\author{Richard A. Battye and Dominic G. Viatic}
\address{Jodrell Bank Centre for Astrophysics, Department of Physics and Astronomy,  University of Manchester, 
Manchester M13 9PL, U.K.}

\label{firstpage}

\date{\today}

\begin{abstract} 
Using a toy model for the interactions between a defect-forming field and the photon field where the photon becomes massive in the defect core (motivated by recent work on defects in the 2HDM), we study the impact on photon propagation in the background of the defect. We find that, when the photon frequency (in natural units) is much lower than the symmetry breaking scale, domain walls reflect most of an incoming photon signal leading to potential interesting astrophysical signals. We also adapt the calculations for vortices and monopoles. We find that the case of strings is very similar to the standard case for massive scalar particles, but in the case of monopoles the cross-section is proportional to the geometrical area of the monopole.
\end{abstract}



\maketitle

\section{Introduction}

Topological defects are expected to form via the Kibble Mechanism during the spontaneous breakdown of exact symmetries in gauge theories where there is  non-trivial homotopy of the vacuum manifold~\cite{Kibble:1976sj}. These relics of the high energy era have been postulated to lead to a very wide range of physical phenomena in cosmology and astrophysics (see the many excellent reviews of the subject over the years~\cite{Shellard1994,Hindmarsh:1994re,Vachaspati:2015cma}). The gravitational effects of topological defects are thought to be universal and not dependent on the microphysical details of the model in which the defects exist. For this reason, they are the most commonly studied phenomenological consequences and there are strong constraints. In particular, the mass per unit length of cosmic strings, $\mu$, is constrained to be $G\mu\lesssim 10^{-7}$, where $G$ is the gravitational constant and we have set the speed of light $c=1$ in natural units which we will use throughout, from observations of the Cosmic Microwave Background (CMB)~\cite{Planck:2011ai} and there is a constraint in the range $G\mu\lesssim 10^{-7}-10^{-11}$ from the absence of timing residuals in pulsar timing (see, for example, \cite{Sanidas:2012ee}). 

Predictions for generic particle interactions with topological defects are more difficult to make and, as a result, are less standard. It is generally accepted that there will be a friction dominated era just after the formation of the defects when density of particles in the thermal bath is sufficiently high to inhibit their motion~\cite{Everett:1981nj,Vilenkin:1991zk}. Once the density of the Universe reduces, cosmic strings and domain walls are expected to move under their tension and evolve toward a scaling solution, while monopoles just evolve like a background of point particles, and eventually come to dominate the evolution of the Universe. Other than this friction dominated era, particle interactions are typically, but not always, ignored.

It has been suggested that a wide class of topological defect models will be superconducting~\cite{Witten:1984eb,Hodges:1988qg,Davis:1995kk,Brandenberger:1996zp}. This effect results from the interactions of the defect forming field with one associated with an unbroken symmetry which allows Noether charge to condense and currents to flow. It is expected that this can lead to a range of interesting phenomenological consequences. The term superconducting is used very generally in this context to refer to a defect in which there is charge and current, but in this work we will use it to mean a topological defect where the photon mass is non-zero in the core of the superconducting defect; that is, the charge and current are related to those of standard electromagnetism. This is the conventional definition of ``superconducting'' used in the literature in the field of condensed matter

Our study is motivated by the observation~\cite{Battye:2020sxy} in simulations of the formation and evolution of domain walls in the Two Higgs Doublet Model (2HDM)~\cite{Lee1973}. This model is a simple and well-studied extension of the Standard Model (SM) of Particle Physics~\cite{Pilaftsis1999,Branco2012,Ivanov2017}. As shown in \cite{Brawn2011} it can, under certain parameter choices, have accidental global symmetries which can lead to topological defects. These can be of two varieties: Higgs Family (HF) and Charge-Parity (CP). In what follows we will concentrate on those with HF symmetries which can be $Z_2$, $U(1)$ or $SU(2)$ symmetric leading to domain walls, vortices (cosmic strings in 3D) and monopoles, although we believe the basic ideas will also apply to the CP symmetries. What was shown in \cite{Battye:2020sxy} was that the domain walls formed by the breaking of the $Z_2$ are not typically the lowest energy, charge neutral  state, but have a higher energy and violate charge neutrality in the core of the wall. This violation of charge neutrality means that the photon acquires a mass in the core; something which we believe is likely to also be the case in simulations of strings and monopoles formed by the breaking of $U(1)$ and $SU(2)$ HF symmetries, respectively.

In this short paper we will study the photon interactions with topological defects in a toy field theoretic model where the photon has a mass in the core of the defect as seen the 2HDM. This model has been constructed specifically to allow our calculations and is meant to be {\it representative} of a more complicated model, such as in the 2HDM, which is well-defined, rather than being a model which in isolation exists in nature. The key feature it will have is that there is a simple interaction between the scalar field and the gauge field associated with photon that leads to a mass for the photon in the core of the defect that we will be able to use to make predictions of how a photon will interact with the defect. We will see that there are interesting phenomenological consequences, particularly in the case of domain walls. We will consider more detailed, but consequentially more complicated calculations, within the 2HDM in a future publication.

\section{Toy model}
The model which we will consider in this paper has Lagrangian density
\be
{\cal L}={1\over 2}|\partial_\mu\mathbf{\sigma}|^2-{1\over 4}\lambda\left(|\mathbf{\sigma}|^2-\eta^2\right)^2-{1\over 4}F_{\mu\nu}F^{\mu\nu}-{\epsilon\over 2\eta^2}\left(\mathbf{\sigma}|^2-\eta^2\right)^2A^{\nu}A_{\nu}\,,
\ee
where $\mathbf{\sigma}=(\sigma_1, ..,\sigma_n)$ for $n\le 3$ and $A_{\mu}$ is the gauge field representing the photon field. $\lambda$ and $\epsilon$ are dimensionless coupling constants  and the symmetry breaking scale, $\eta$, has dimensions of energy. The final term has been chosen so that the mass of the photon is always positive. The equations of motion are 
\bea
\Box\sigma_i+\lambda\sigma_i\left(|\mathbf{\sigma}|^2-\eta^2\right)+{2\epsilon\over\eta^2}\sigma_i(|\sigma|^2-\eta^2)A^{\mu}A_{\mu}&=&0\,,\cr
\Box A_{\nu}-\partial_\nu(\partial^\mu A_{\mu})+{\epsilon\over\eta^2}\left(|\mathbf{\sigma}|^2-\eta^2\right)^2A_{\nu}&=&0\,.
\eea
These equations admit a solution with $A_{\mu}\equiv 0$. Therefore, we will consider solutions of the form $\sigma_i({\bf x},t)={\bar\sigma}_i({\bf x})+\delta\sigma_i({\bf x},t)$ and $A_{\mu}({\bf x},t)=\delta A_{\mu}({\bf x},t)$. At zeroth order, we find that $-\nabla^2{\bar\sigma}_i+\lambda{\bar\sigma}_i\left(|\mathbf{\bar{\sigma}}|^2-\eta^2\right)=0$, and at first order, we have that 
\bea
\Box{\delta\sigma}_i+\lambda\left[\left(|{\bar{\sigma}}|^2-\eta^2\right)\delta_{ij}+2{\bar\sigma}_i{\bar\sigma}_j\right]\delta\sigma_j&=&0\,,\cr
\Box \delta A_{\nu}-\partial_\nu(\partial^\mu\delta A_\mu)+{\epsilon\over\eta^2}\left(|{\bar\sigma}|^2-\eta^2\right)^2\delta A_{\nu}&=&0\,.
\eea
The first of these perturbative equations is the standard equation for the massive excitations and, when $n>1$, massless excitations. This is well understood and we will not be concerned with this here. The second equation is for the evolution of the gauge (photon) field in the background ${\bar\sigma}_i({\bf x})$.

The zeroth-order equation has topological solutions in spatial dimension $n$, corresponding to domain walls ($n=1$), vortices ($n=2$) and monopoles ($n=3$). The domain wall solution is ${\bar\sigma}=\sigma_1=\eta\tanh(x/\Delta)$ where $\Delta=(\lambda/2)^{-1/2}\eta^{-1}$ is the width of the domain walls. The solution has $|{\bar\sigma}|=\eta$ at $x=\pm\infty$ and $|{\bar\sigma}|=0$ at $x=0$. The vortex and monopole solution in 2D and 3D, respectively, are given by ${\bar\sigma}_i=\eta f(r\eta\sqrt{\lambda}){\hat{r}}_i$ where ${\hat r}_i$ is the unit position vector in the appropriate dimension. The profile function satisfies 
\be
{d^2f\over dr^2}+{n-1\over r}{df\over dr}+(n-1)f+f(f^2-1)=0\,,
\ee
and has boundary conditions $f(0)=0$ and $f(r)\rightarrow 1$ as $r\rightarrow\infty$. Note that the mass per unit length of the string solution here is logarithmically divergent and that of the monopole is linearly divergent; both would be cut-off\footnote{The need to impose these cut-offs is the consequence of the broken symmetry being global; a choice made to simplify the calculations within this toy model. If it were a local, gauge symmetry then the infra-red divergences would be removed, while the ultra-violet divergence in the string core would remain and $\Delta$ would depend on the gauge couplings.} in a realistic context a large distances by the inter-defect distance, $\Lambda$, and at small distances, in the case of the vortex, by the string core distance, $\Delta$. The important feature of the discussion in the present context is that the photon mass, $m_{\gamma}^2=\epsilon(|\mathbf{\bar\sigma}|^2-\eta^2)^2/\eta^2$, is zero at  $\infty$ and is $m_\gamma^*=\epsilon^{1/2}\eta$ in the defect core.

It is important in the calculations which follow to be able to impose the gauge condition $\partial_\mu\delta A^{\mu}=0$ in order for the photon field to have the correct properties. If the mass of the photon is constant in space, then this condition can be derived as a consequence of antisymmetric electromagnetic field tensor by differentiating the equations of motion, even though the Lagrangian does not have gauge invariance. Re-establishing gauge invariance in a massive photon theory can be achieved by the so-called ``Stuckelberg Trick" (see, for example,  \cite{Ruegg:2003ps}) which involves the introduction of extra, auxiliary fields. However, we have already pointed out that our toy model is to be thought of as representing aspects of the behaviour of a more complete model, such as, the 2HDM, which will naturally have gauge invariance and the correct properties of the photon field. For this reason we will impose the condition on the solutions without it being naturally part of the toy model, presuming that whatever the global model is, allows this.

\section{Scattering solutions}

\subsection{One dimension}

Using the domain wall solution for ${\bar\sigma}$ we can write the equation for the photon field as 
\be
\Box\delta A_{\nu}+\epsilon\eta^2\sech^4\left({z\over\Delta}\right)\delta A_{\nu}=0\,.
\ee
If we assume that the width of the wall is small compared to the wavelength of the incoming photon, then we can replace $\sech^4(z/\Delta)\rightarrow 4\Delta\delta(z)/3$ with the coefficient defined to make the integrals over the profile the same on both sides. If we now make the ansatz $\delta A_{\mu}=n_{\mu}\exp[i\omega t]{\widetilde{\delta A}}(z)$, where $n_{\mu}$ is a constant polarisation vector, and hence we are left with 
\be
-{d^2\over dz^2}{\widetilde{\delta A}}+\left[{4\epsilon\over 3}\left({2\over \lambda}\right)^{1/2}\eta\delta(z)-\omega^2\right]{\widetilde{\delta A}}=0\,,
\ee
which is a Schr\"odinger equation for $\delta$-function potential. If we now consider a solutions to this equation with boundary conditions 
\be
{\widetilde{\delta A}}(z)=\bigg\{\begin{aligned}e^{i\omega z}+{\cal R}e^{-i\omega z}  \qquad z<0\,,\\  {\cal T}e^{i\omega z} \qquad z>0\,,\end{aligned}
\ee
then we deduce that the transmission probability is 
\be
|{\cal T}|^2=\left(1+{8\epsilon^2\eta^2\over 9\lambda\omega^2}\right)^{-1}\,,
\label{trans}
\ee
and therefore in the low frequency limit we find a suppression of the transmission of photons by the wall which is $\propto\lambda\omega^2/(\epsilon^2\eta^2)$. Most astrophysically generated photons are in this regime and, therefore, we see that the walls will reflect most of their photons with significant observational consequences that we will discuss later.

\subsection{Two and three dimensions}

The cases of vortices in 2D ($n=2$) and monopoles in 3D ($n=3$) can be treated, at least initially, together. In what follows we will follow closely the textbook treatment for the interactions of a string with scalar massive particles presented in Chapter 8 of \cite{Shellard1994} which in turn relies on \cite{Everett:1981nj}. We will first define $({\bar\sigma}^2(r)-\eta^2)^2\equiv\eta^4F(r\eta\sqrt{\lambda})$ where $F(0)=1$ and $F(r)\rightarrow 0$ as $r\rightarrow\infty$ and make the same substitution, $\delta A_{\nu}=n_{\nu}\exp[i\omega t]{\widetilde{\delta A}}({\bf x})$ where ${\bf x}=(x_1,..,x_n)$, yielding the equation of motion 
\be
\left(\nabla^2+\omega^2\right){\widetilde{\delta A}}=\epsilon \eta^2F{\widetilde{\delta A}}\,.
\ee

As with case of the domain wall we replace the profile function $F$ with a delta function $F(r\eta\sqrt{\lambda})=I_n/(\eta\sqrt{\lambda})^{n}\delta^{[n]}({\bf x})$ where $I_n=\Omega^{[n-1]}\int_0^{\infty} x^{n-1}F(x)dx$ and $\Omega^{[q]}$ is the area of $S^{q}$, that is, $\Omega^{[1]}=2\pi$ and $\Omega^{[2]}=2\pi$. In this approximation we have that 
\be
\left(\nabla^2+\omega^2\right){\widetilde{\delta A}}=\kappa{\widetilde{\delta A}}(0)\delta^{[n]}({\bf x})\,,
\ee
where $\kappa=\epsilon\eta^{2-n}\lambda^{-n/2}I_n$. As described in \cite{Shellard1994}, one can search for a scattering solution
\be
{\widetilde{\delta A}}({\bf x})=e^{i\boldsymbol{\omega}\cdot{\bf x}}-\kappa{\widetilde{\delta A}}(0)G({\bf x})\,,
\ee
where 
\be 
G({\bf x})={1\over (2\pi)^n}\int d^n{\bf p}{e^{i{\bf p}\cdot{\bf x}}\over p^2-\omega^2-i\varepsilon}\,,
\ee
is the Green function and $\varepsilon\rightarrow 0^{+}$. We note that ${\widetilde{\delta A}}(0)=[1+\kappa G(0)]^{-1}$. At this stage it is best to discuss the two cases $n=2$ and $n=3$ individually.

\medskip
\noindent (i) $n=2$. For $|{\bf x}|=r\gg\omega^{-1}$ the 2D Green function is asymptotically $G({\bf x})\approx\left({i\over 8\pi \omega r}\right)^{1\over 2}e^{i\omega r}$, but it is logarithmically divergent as $r\rightarrow 0$ and hence one is forced to introduce a cut-off around the scale of the string core, $\Delta$. One finds that $G(0)\approx-{1\over 2\pi}\log(\omega\Delta)$ and hence the differential cross-section per unit length of string is 
\be
{d\sigma_{\rm 2D}\over d\theta}={\kappa^2\over 8\pi \omega\left[1-{\kappa\over 2\pi}\log(\omega\Delta)\right]^2}\approx\bigg\{\begin{aligned}
&{\kappa^2\over 8\pi\omega}\,,&\cr &{\pi\over 2\omega[\log(\omega\Delta)]^2}\,,&\end{aligned}
\ee
where $\kappa^2\propto ({\epsilon/\lambda})^2\propto \left(m_\gamma^*/ m_\phi\right)^4$ and $m^2_\phi=\lambda\eta^2$ is the mass of the string forming field. The approximations above are made when the logarithmic term is sub-dominant (upper) and dominates (lower). In both cases, apart from the weak logarithmic dependence, we have that the cross-section per unit length is $\propto\omega^{-1}$. The case where the logarithm dominates could have been anticipated by the standard calculation for a scalar field interacting with the string forming fields \cite{Everett:1981nj,Shellard1994}.  In that case the result is independent of the photon mass and only weakly depends on the symmetry breaking scale (and hence the width of the string). The key feature that we have argued for is that, this formula also applies when the string is superconducting in the traditional sense that the photon has mass in the string core. This result is a consequence of calculations in ref.~\cite{Witten:1984eb}.

\medskip 
\noindent (ii) $n=3$. The 3D Green function is $G({\bf x})=e^{i\omega r}/(4\pi r)$ which is divergent, but this time with a pole singularity $G(0)\approx (2\pi^2\Delta)^{-1}$. The differential cross-section is given by 
\be
{d\sigma_{\rm 3D}\over d\Omega}={\kappa^2\over 16\pi^2\left[1+{\kappa\over 2\pi^2\Delta}\right]^2}\,.
\ee
Note that $\kappa/\Delta\propto\epsilon/\lambda\propto(m_\gamma^*/m_\phi)^2$ and the result is independent of the frequency. Hence, the differential cross section is  always proportional to the geometrical cross section of the monopole, that is, $ \propto \Delta^2$.
When the pole dominates, we have that total cross section is $\sigma_{\rm tot}\approx\pi^3\Delta^2$ . The mean free path for superconducting monopoles moving the Cosmic Microwave Background (CMB) would be  $\lambda_{\rm F}\sim (n\sigma_{\rm tot})^{-1}\sim \eta^2/T^3$ where the number density of photons $n\sim T^3$ and at the present day it would be that  $\lambda_{\rm F}\sim 10^6\,{\rm Mpc}\gg H_0^{-1}$ if $\eta\sim 1\,{\rm TeV}$ implying that they would have decoupled in the late-universe.

\section{Discussion}

The key new result of this paper is that a superconducting domain wall would lead to photons being reflected rather than being transmitted\footnote{We note that spontaneously broken discrete symmetries are already strongly constrained \cite{Zeldovich:1974uw} and, in particular, strong constraints exist on the parameters of a $Z_2$ Symmetric 2HDM model which produces walls \cite{Battye:2020jeu}, but it was shown that it is still interesting to study these interactions in the context of approximate symmetries.}. This would have interesting phenomenological consequences: CMB photons would not be transmitted through the walls since $\omega\ll \eta$; the fact that we have observed and measured the properties of the CMB to exquisite detail suggests that the level of attenuation predicted in this calculations has not taken place unless $\epsilon$ is very small.


In addition to this we have derived qualitative results that one would expect for strings and monopoles. The string result in the case where the logarithm is dominant could have been anticipated from textbook results \cite{Shellard1994} suggesting that there are no specific phenomenological consequences, while that for monopoles is simply proportional to the geometrical area, $\propto\Delta^2$ independent of frequency. This could be interesting since it is standard to assume that the cross-section for monopole scattering with charged particles is $\propto T^{-2}$ since this leads to a different dependence for the mean-free path, $\lambda_{\rm F}\propto T^{-1}$. The motivating work for this study \cite{Battye:2020sxy} concentrated on the case of domain walls, but our preliminary indications are that global monopoles formed in the 2HDM with $SU(2)_{HF}$ symmetry also have a charge violating vacuum in their core, leading to a non-zero photon mass. Such a model is difficult to imagine phenomenologically, but we would expect similar effects in the case of strings and our results provide motivation for further investigations of this case.

An important caveat that we should highlight is that these calculations would not apply in the case where the incoming photon frequency is comparable (in natural units) to the energy scale associated with the symmetry breaking. In this case, one would expect much more complicated interactions which would involve interaction with the other field quanta which the string comprises, such as the gauge bosons. A more detailed study of this would need to be done in the context of a particular model which we are presently engaged.

This work was motivated by the 2HDM. However, we believe that the basic results and concepts apply to superconducting defects more generally. There has been quite a bit of interest recent on superconducting topological defects in a wide range of contexts \cite{Abe:2020ure,Fukuda:2020kym}, way beyond the confines of the 2HDM. Our results suggest that such defects might have interesting phenomenological consequences should they survive to the present day and suggest that it is worth considering defect interactions beyond the standard lore of constraining them via their gravitational effects which are typically very weak for intermediate- and weak-scale symmetry breaking.
 
\section*{Acknowledgements}
\noindent It is a pleasure to thank Apostolos Pilaftsis for his collaboration on related work and for helpful comments on this paper. We have also benefitted from comments from Tanmay Vachaspati.
\appendix

\bibliographystyle{apsrev4-1}
\bibliography{photonint}

\label{lastpage}

\end{document}